\documentstyle[11pt,paspconf,164page,164def,psfig]{article}
\begin{document}

\def\simleq{\stackrel{<}{\scriptstyle\sim}}
\def\simgeq{\stackrel{>}{\scriptstyle\sim}}

\title{Variability of the Radio Nucleus of the Galaxy M81}

\index{AGN:variability|(}
\index{M81|(}
\label{biete}
\markboth{Bietenholz, Bartel \& Rupen}{Variability of the Radio Nucleus of M81}

\author{M. F. Bietenholz, N. Bartel} \affil{Department of Physics and
Astronomy, York University, North York, M3J~1P3, Ontario, Canada}

\author{M. P. Rupen}
\affil{National Radio Astronomy Observatory, Socorro, New Mexico 87801, USA}

\begin{abstract} We report on VLA and VLBI observations of the nucleus
of the nearby spiral galaxy M81.  The VLA observations show the flux
density of the nucleus to be variable by 50\%. The VLBI observations
indicate that the structure of the nucleus of M81 is somewhat variable
on timescales of weeks.
\end{abstract}

\keywords{AGN: individual}

\section{Introduction}

M81, at a distance of 4~Mpc, is the nearest galaxy with an active
galactic nucleus, with the possibly exception of Cen A.  The nucleus is
exceptionally compact: 700~AU $\times$ 300~AU at 22~GHz, with the size
being proportional to $\nu^{-0.8}$ (Bietenholz \etal\ 1996; Bartel
\etal\ 1995; Bartel \etal\ 1982; see also Kellermann \etal\
1976). Here we present further results from new VLA and VLBI
observations of the nucleus of M81.

\section{VLA Flux Density Monitoring}

\begin{figure}[h]
\cl{\psfig{figure=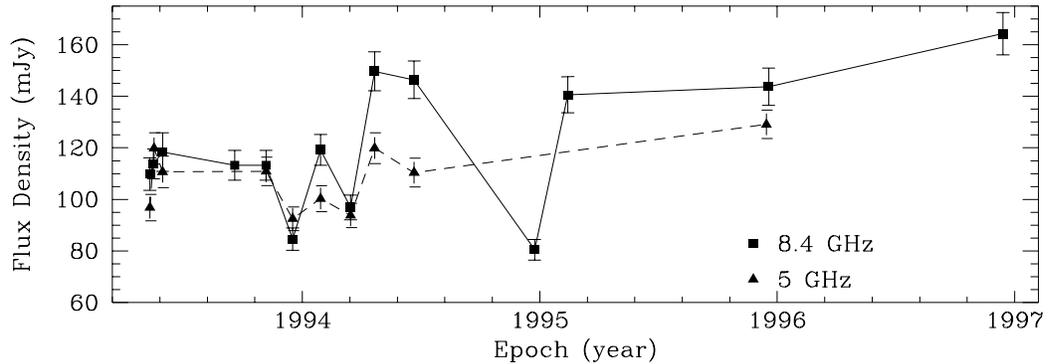,width=1.1\textwidth,angle=-90}}
\caption{VLA Flux Density Measurements\label{bietef1}}
\end{figure}

The nucleus of M81 was used as a reference source for the continuing
program of VLBI monitoring of SN1993J (see Rupen~\etal\, these
proceedings p.~\ref{rupen}).  VLA data were taken simultaneously with
the VLBI runs.  Flux densities were derived from images, and the
results are shown in Figure~\ref{bietef1}.  The systematic uncertainty
in the flux calibration (5\%) dominates internal uncertainties.  The
flux density of M81 varies by over a factor of 2 at 8.4~GHz (see also
van Dyk \& Ho, these proceedings p.~\ref{vanho}).  At this frequency,
the mean observed flux density was 120~mJy with a standard deviation
of 22~mJy or 19\%.  The flux densities at 5 and 8.4~GHz seem well
correlated, although the radio spectral index ($\alpha$, where $S
\propto\nu^\alpha$) is not constant within our errors:
$\alpha$ varies by up to $3.2\sigma$.

\psrotatefirst

\section{VLBI Imaging and Modelfitting \label{bietes3}}

In our VLBI observations, the nucleus of M81 is only slightly resolved
and somewhat asymmetrical (in agreement with Bietenholz \etal\
1996). In Table~\ref{bietet1}, we give the results of fitting simple
geometrical models to fully calibrated {\em u-v} data.

\index{M81!structure variability}
\begin{table}[t]
\caption{Summary of the VLBI Model-Fitting Results at 8.4~GHz \label{bietet1}}
{\footnotesize
\begin{center}
\begin{tabular}{c @{\hskip 17pt} c c c c @{\hskip 40pt} 
                      c c c @{\hskip 11pt} r @{\hskip 8pt} c c}
\hline\hline
\noalign{\vspace{1pt}}
        & &\multicolumn{3}{c}{One Component Fit} 
                       &\multicolumn{6}{c}{Two Component Fit} \\

        & &\multicolumn{3}{c}{Elliptical Gaussian}
                &\multicolumn{3}{c}{Elliptical Gaussian}
                &\multicolumn{3}{c}{Point Source}\\ \hline

\noalign{\vspace{1pt}}
Date   & $S$  &\multicolumn{2}{c}{Axes}& P.A.
              &\multicolumn{2}{c}{Axes}& P.A. & \% of $S$
              & $r$ & $\theta$ \\

        &    & Maj.  & Min.  &     & Maj.  & Min. &     &  \\ \hline
\noalign{\vspace{1pt}}
{\tiny dd/mm/yy}    
             & mJy & mas   & mas   &\deg & mas   & mas  & \deg & & mas & \deg \\ 
                \hline
16/05/93     & 114 & 0.44 & 0.17 & 58 & 0.40 & 0.18 & 57 & 5~  & 0.57 & 74 \\
16/12/93     & \hspace{4pt}85 
                   & 0.44 & 0.21 & 50 & 0.42 & 0.17 & 46 & 5~  & 0.61 & 81 \\
21/06/94     & 146 & 0.61 & 0.18 & 45 & 0.48 & 0.17 & 41 & 14~ & 0.58 & 51 \\
12/02/95     & 141 & 0.44 & 0.19 & 50 & 0.44 & 0.13 & 50 & 6~  & 0.87 & 45 \\
18/12/95     & 144 & 0.43 & 0.23 & 49 & 0.40 & 0.17 & 61 & 13~ & 0.62 & 57 \\
\hline
\noalign{\vspace{1pt}}
\multicolumn{5}{r}{approximate standard errors}
                                        & 0.05 & 0.05 &
                                             \hspace{4pt}5 &  2~ & 0.10 & 10 \\
\hline
\end{tabular}
\end{center}
~~Notes: $S$ are flux densities from VLA observations at 8.4~GHz; 
axis sizes are FWHM.
}
\end{table}

We used the AIPS program OMFIT to simultaneously fit model parameters
and determine the complex antenna gains.  In the simplest case of
fitting a single elliptical Gaussian, the size varies by 40\%. The
data, however, are not well described by such a model: in all cases,
the fit is significantly improved if we fit, in addition to the central
elliptical Gaussian source, a weaker point source at relative
position $(r, \theta)$.  Though our resolution is inadequate to determine the
detailed source structure, the data clearly demand structure more
complicated than a single elliptical Gaussian.  Note that our
conservative estimates of the uncertainties take into account any
contribution from the fit time-variable antenna gains (a more complete
description will be published elsewhere).

In summary, we find that there is structure on scales of $\sim
0.5$~mas in the nucleus of M81. 
It can be described as an elongated core, with a north-east south-west
orientation, and a component of lower flux density to the north-east
thereof.  This latter component, in particular, appears to be variable
on timescales of weeks.

\index{AGN:variability|)}
\index{M81|)}
\footnotesize\acknowledgments

Research at York University was partly supported by NSERC.
\NRAOcredit

\end{document}